\pdfoutput=1                       
                       \documentclass[9pt,twocolumn,twoside]{pnas-new}

\templatetype{pnasresearcharticle}

\usepackage{mathtools}

\begin{document}

\title{Spatially resolved random telegraph fluctuations of a single trap at the Si/SiO\textsubscript{2} interface}

\author[a]{Megan Cowie}
\author[b]{Procopios C. Constantinou}
\author[b,c]{Neil J. Curson}
\author[b,c]{Taylor J.Z. Stock}
\author[a]{Peter Gr\"{u}tter}

\affil[a]{Department of Physics, McGill University, Montr\'eal, Qu\'ebec, Canada, H3A 2T8}
\affil[b]{London Centre for Nanotechnology, University College London, London, United Kingdom, WC1H 0AH}
\affil[c]{Department of Electronic and Electrical Engineering, University College London, London, United Kingdom, WC1E 7JE}

\leadauthor{Cowie}

\significancestatement{Low-frequency noise due to two-level fluctuations inhibits the reliability and performance of nanoscale semiconductor devices, and challenges the scaling of emerging spin-based quantum sensors and computers. Here, we measure temporal two-state fluctuations of individual defects at the Si/SiO\textsubscript{2} interface with nanometer spatial resolution using atomic force microscopy. When measured as an ensemble, the observed defects have a $1/f$ power spectral trend at low frequencies. The presented method and insights provide a more detailed understanding of the origins of $1/f$ noise in silicon-based classical and quantum devices, and could be used to develop processing techniques to reduce two-state fluctuations associated with defects.}

\authorcontributions{T.S., P.C., and N.C. fabricated the sample. P.G. initiated the research. M.C. performed the measurements and analysis. P.G., N.C., and T.S. secured funding. M.C., P.G., and T.S. wrote the paper.}
\authordeclaration{The authors declare no competing interests.}
\correspondingauthor{\textsuperscript{2}To whom correspondence should be addressed. E-mail: megan.cowie@mail.mcgill.ca}


\begin{abstract}
We use electrostatic force microscopy to spatially resolve random telegraph noise at the Si/SiO\textsubscript{2} interface. Our measurements demonstrate that two-state fluctuations are localized at interfacial traps, with bias-dependent rates and amplitudes. These two-level systems lead to  correlated carrier number and mobility fluctuations with a range of characteristic timescales; taken together as an ensemble, they give rise to a $1/f$ power spectral trend. Such individual defect fluctuations at the Si/SiO\textsubscript{2} interface impair the performance and reliability of nanoscale semiconductor devices, and will be a significant source of noise in semiconductor-based quantum sensors and computers. The fluctuations measured here are associated with a four-fold competition of rates, including slow two-state switching on the order of seconds and, in one state, fast switching on the order of nanoseconds which is associated with energy loss.
\end{abstract}

\dates{This manuscript was compiled on \today}
\doi{\url{www.pnas.org/cgi/doi/10.1073/pnas.XXXXXXXXXX}}

\maketitle
\thispagestyle{firststyle}
\ifthenelse{\boolean{shortarticle}}{\ifthenelse{\boolean{singlecolumn}}{\abscontentformatted}{\abscontent}}{}

\firstpage[3]{4}


Low-frequency $1/f$ noise in silicon-based field-effect devices is widely attributed to random fluctuations in the charge state occupancy and/or structure of traps at the Si/SiO\textsubscript{2} interface. This random telegraph noise (RTN) compromises circuit performance and reliability, and is increasingly detrimental as device areas decrease in size, in highly scaled devices. RTN in nanoscale metal-oxide-semiconductor field effect transistors (MOSFETs), for example, can introduce current variations that are comparable to the channel signal\cite{Simoen2011,Grasser2020}. RTN also decreases the read margin for random access memory, which limits device stability and scaling\cite{Zahoor2020,Grasser2020}. Emerging semiconductor-based quantum sensors and computers are also prone to RTN. Spin qubits are realized in silicon either as gate-defined quantum dots or as buried dopant atoms positioned nanometers beneath a Si/SiO\textsubscript{2} interface\cite{Morton2011}. In both architectures, RTN in the near-interfacial silicon electronic bath, due to Si/SiO\textsubscript{2} trap fluctuations, will significantly limit qubit coherence and control in noisy intermediate-scale quantum (NISQ) silicon devices\cite{Preskill2018}. 

A more robust understanding of RTN is needed to establish silicon nanofabrication methodologies that limit device noise, and will aid in the development of error correction algorithms for upcoming NISQ technologies. However, despite more than 50 years of research investigating $1/f$ noise in silicon, there is still significant debate regarding what kinds of defects predominantly contribute to its origin. The two defect classes typically associated with RTN in silicon-based devices are interfacial charge traps, which are silicon dangling bonds located exactly at the Si/SiO\textsubscript{2} interface, and oxide traps, which are oxygen vacancies positioned deeper within the oxide layer\cite{Engstrom2014,Fleetwood2023,Grasser2020,Simoen2016}.

In field-effect silicon devices, RTN is commonly attributed either to fluctuations in carrier number, which is associated with a variable interfacial capacitance as the defect charge state switches, or to fluctuations in the carrier mobility due to variable trap scattering\cite{Grasser2020,Kirton1989,Fleetwood2023}. There is increasing consensus that number and mobility fluctuations are correlated and both contribute to device noise, but the exact noise mechanism remains unclear. This is in part because oxide traps and interface traps contribute very differently to carrier number and mobility fluctuation models. The rate of carrier exchange with oxide traps is consistent with commonly observed slow ($\sim \mathrm{ms-s}$) RTN timescales. However, oxide traps are too far removed, spatially, from the Si/SiO\textsubscript{2} interface to participate appreciably in scattering with the silicon surface charge density, and therefore are not expected to be associated with significant mobility fluctuations\cite{Fleetwood2023}. Interface traps, on the other hand, due to their close proximity to the silicon surface, are effective scattering sites and so could give rise to mobility fluctuations, but carrier exchange with interface traps is expected to be too fast to account for slow RTN timescales\cite{Contaret2006,Srinivasan2005,Haartman2007}. 

Part of the challenge in pinpointing the dominant origin of RTN is that the constituent random telegraph signals (RTSs, due to individual two-state fluctuators) and $1/f$ noise (due to an ensemble of two-state fluctuators) are commonly studied by measuring the drain current or voltage in MOSFET devices. In this measurement scheme, distinguishing RTSs associated with individual trap fluctuations requires small and pristine MOSFET devices with few RTS sources. Even then, the exact position of the trap in the MOSFET channel is unknown. Furthermore, both carrier number and mobility fluctuations manifest as MOSFET current switching, which makes it difficult to disentangle these two correlated yet potentially competing noise mechanisms. 

In this work, we demonstrate spatially resolved RTN at an n-type Si/SiO\textsubscript{2} interface using frequency modulated atomic force microscopy (fm-AFM)\cite{Albrecht1991,Garcia2002} at room temperature in ultra-high vacuum ($\sim\mathrm{10^{-10}~mbar}$). We measure nanoscale spatial heterogeneities in the noise landscape at the silicon surface, and attribute this to localization of two-state fluctuators at charge trap sites. In these measurements, fluctuations in the interfacial capacitance and carrier scattering rates, which are associated with the local carrier number and mobility, respectively, are measured simultaneously. In this way, we detect correlated fluctuations, with a range of characteristic timescales, at individual trap sites with bias-dependent rates and amplitudes
consistent with  scattering by donor-like
interface traps\cite{Cowie2023}.

\section*{Results}\label{Sec:Results}
The fm-AFM tip-sample junction resembles a metal-insulator-semiconductor (MIS) capacitor, where the tip (which has an applied bias $V_g$) is metallic and the n-type silicon sample (grounded) is semiconducting, and there is an insulating gap comprised of vacuum ($\mathrm{\sim10~nm}$) and SiO\textsubscript{2} ($\mathrm{1~nm}$) between them\cite{Cowie2022,Cowie2023,Feenstra2006,Johnson2011,Johnson2009,DiBernardo2022,Wang2015,Winslow2011}. As the fm-AFM cantilever oscillates above the silicon surface, the surface charge density (i.e. surface potential) varies in time -- that is, the effect of the oscillating cantilever is similar to the application of an AC bias at the cantilever resonance frequency, $\mathrm{300~kHz}$. This leads to a time-varying tip-sample force ($F_{ts}$) with an in-phase contribution, with respect to the tip-sample separation ($z_{ins}$), as well as an out of phase contribution. The in-phase force is related to the interfacial capacitance\cite{Hudlet1995,Cowie2022}, and manifests as a shift in the cantilever resonance frequency ($\Delta f$). The out-of-phase force, which is non-zero if the surface charge re-organization is non-instantaneous, is related to the equivalent series resistance, or Ohmic energy dissipation due to scattering of mobile carriers near the surface as they reposition over every oscillation cycle. This out-of-phase force manifests as an increase in the amplitude of the cantilever driving force ($F_d$)\cite{Cowie2023}. Thus, we expect any RTN due to capacitance fluctuations to manifest in the fm-AFM $\Delta f$ channel, and RTN due to fluctuating trap structures and scattering centres to manifest in the $F_d$ channel. 


\begin{figure}[!t]
    \centering
    \includegraphics[width=\linewidth]{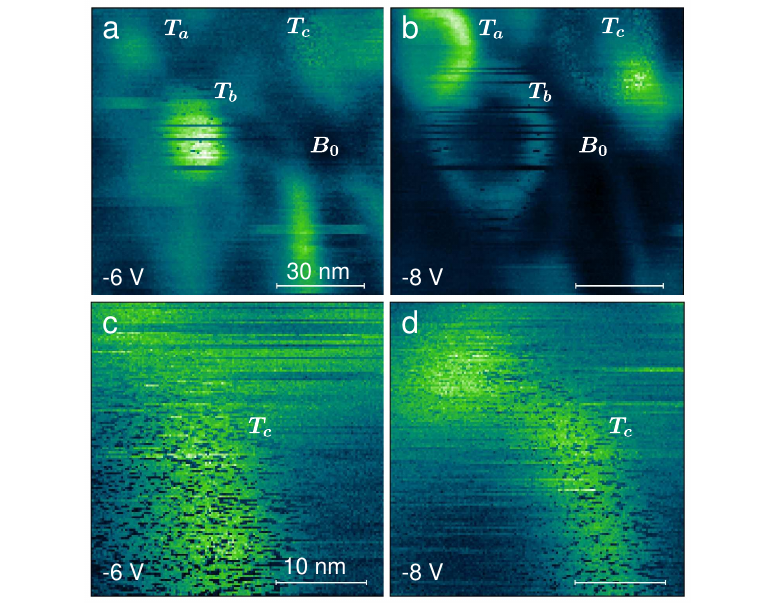}
     \caption{\textbf{fm-AFM driving force ($F_d$) images showing spatially heterogeneous dielectric loss and noise at the SiO\textsubscript{2} interface.} (a,b) Multipass image showing $F_d$ at variable bias (indicated). Three defects labelled $T_a$, $T_b$, and $T_c$ are identified, as well as a trap-free background region $B_0$. The bright regions (as compared to $B_0$) correspond to additional unlabelled donor-like traps. (c,d) Multipass image of the $T_c$ trap. The horizontal scan speed for both images was $\sim 5~\mathrm{s/line}$. The colour scale bars are (a)~${F_d=\mathrm{[30:35]~mV}}$, (b)~${F_d=\mathrm{[31:40]~mV}}$, (c)~${F_d=\mathrm{[32:37]~mV}}$, and (d)~${F_d=\mathrm{[33:41]~mV}}$, where bright green corresponds to an increase in $F_d$ associated with increased dielectric loss.}
    \label{fig:RingofNoise}
\end{figure}
\begin{figure}[!b]
    \centering
    \includegraphics[width=\linewidth]{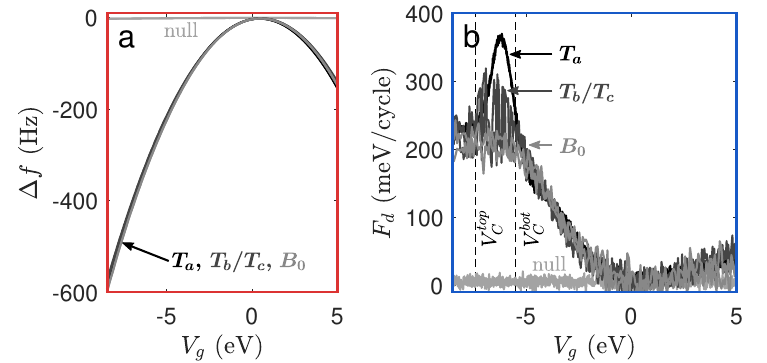}
     \caption{\textbf{fm-AFM bias spectra of the Si/SiO\textsubscript{2} surface.} (a) fm-AFM frequency shift ($\Delta f$) and (b) driving force ($F_d$) at variable bias ($V_g$) measured at the background (light grey, $B_0$), at a $\mathrm{0~Hz}$ trap (black, $T_a$), and at a noisy trap (dark grey, $T_b$ or $T_c$). A null spectrum measured $\mathrm{\sim 1~\mu m}$ above the sample surface (where the tip-sample force $F_{ts}\approx 0$) is also shown (light grey). Between the crossing points (dashed lines, $V_C^{top}$ and $V_{C}^{bot}$), the traps move above and below the Fermi level over every cantilever oscillation\cite{Cowie2023}.}
    \label{fig:NoisySpectra}
\end{figure}

\subsection*{Spatial localization} Figure~\ref{fig:RingofNoise} shows multi-pass images of the fm-AFM driving force $F_d$, where the applied bias $V_g$ is $-6.5~\mathrm{V}$. The measured $F_d$ varies spatially. Specifically, in Figure~\ref{fig:RingofNoise}, three traps (labelled $T_a$, $T_b$, and $T_c$) are identified, as well as a background region labelled $B_0$ (away from any traps). The traps exhibit different noise timescales, ranging from static ($0~\mathrm{Hz}$, $T_a$) to slow ($\sim \mathrm{Hz}$, $T_b$) to fast ($\sim \mathrm{kHz}$, $T_c$). Furthermore, Figure~\ref{fig:RingofNoise}, being a measurement of $F_d$, shows that in the proximity of traps, the mobility at the Si/SiO\textsubscript{2} interface can fluctuate with timescales on the order of $\mathrm{Hz-kHz}$. A more detailed description of this $F_d$ measurement, as well as the slow $F_d$ noise at trap sites, follows.

Figure~\ref{fig:RingofNoise} shows that there is a notable increase in $F_d$ near traps as compared to the background. This corresponds to an increase in scattering loss near trap sites. The loss mechanism is understood by measuring the full bias dependence of $\Delta f$ and $F_d$ at the trap locations as compared to the background, as shown in Figure~\ref{fig:NoisySpectra}. The $F_d$ peaks in Figure~\ref{fig:NoisySpectra}b manifest as rings in Figure~\ref{fig:RingofNoise}; this is due to the spatial localization of the tip, which acts as a gate\cite{Cowie2023}. The bias spectral position of these peaks (rings) is consistent with scattering by a donor-like interface trap, as described in \cite{Cowie2023}: As the fm-AFM cantilever oscillates, the trap level $E_T$ continually shifts with respect to the Fermi level $E_F$. At a particular bias, $V_C^{bot}$, the trap level equals the Fermi level at the bottom of the cantilever oscillation; at $V_C^{top}$, the trap level equals the Fermi level at the top of the oscillation. At biases between these two crossing points, then, the donor-like trap energy moves above and below the Fermi level over every cantilever oscillation cycle. Consequently, within this bias range (i.e. between the dashed lines in Figure~\ref{fig:NoisySpectra}), energy is dissipated by cascade phonon scattering as carriers are repeatedly captured and emitted by the interface trap\cite{Nicollian1967,Lax1960,Wang2007,Piscator2009}, and the measured $F_d$ increases.  This carrier exchange with the donor-like interface trap states occurs faster than the cantilever oscillation period, with timescales on the order of $\mathrm{MHz}$\cite{Cowie2023}. Note that the bias dependence of the trap-free background ($B_0$) $F_d(V_g)$ spectrum in Figure~\ref{fig:NoisySpectra}b is due to the bias-dependent change in surface potential as the cantilever oscillates, as explained in \cite{Cowie2023}. 

\begin{figure*}[!t]
    \centering
    \includegraphics[width=\linewidth]{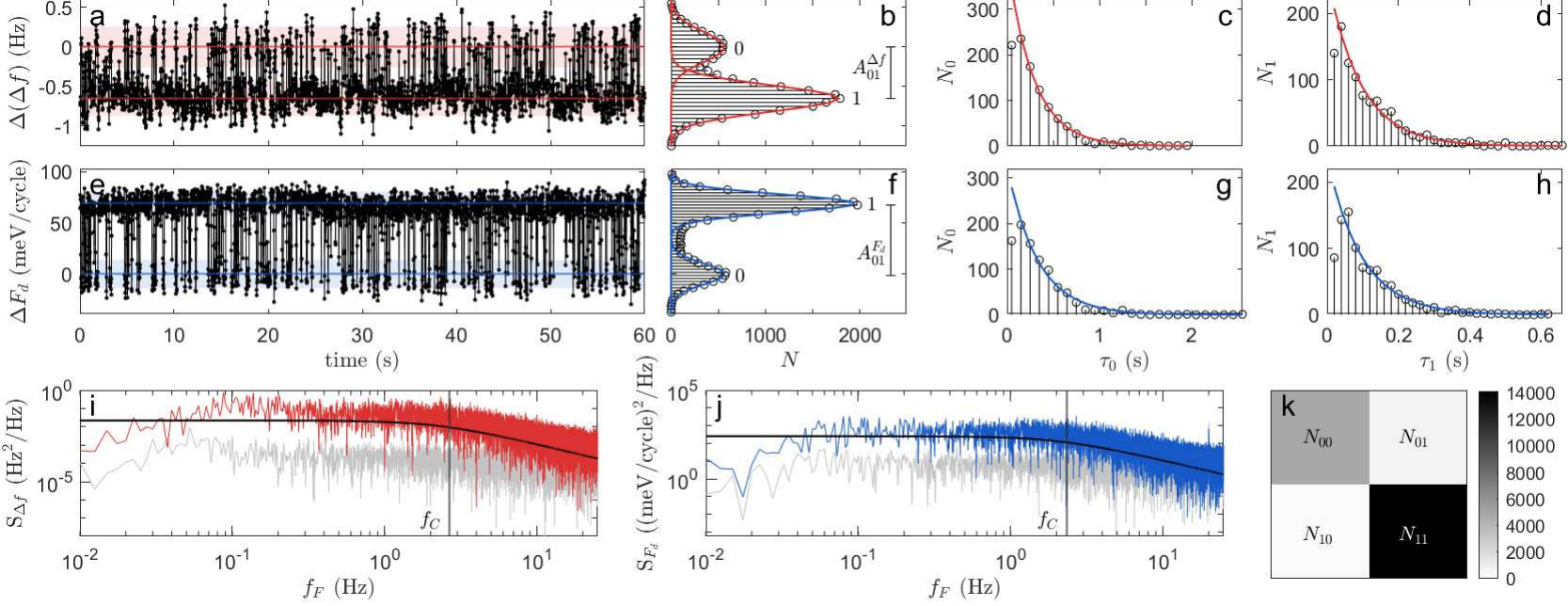}
     \caption{\textbf{Random telegraph signal of an individual interface trap at the Si/SiO\textsubscript{2} interface}. (a,e) Simultaneously measured fm-AFM frequency shift ($\Delta f$) and drive ($F_d$) above a fast ($T_c$) trap at bias $V_g=-6.5~\mathrm{V}$. $60~\mathrm{s}$ of the total $400~\mathrm{s}$ are shown. (b,f) Histograms of the time trace (black) with Gaussian fits (colour) overlaid. The Gaussian mean and full-width half-maximum are shown in (a,e) as coloured lines and shaded regions, respectively. (c-d, g-h) Histograms of the wait time between $0\rightarrow 1$ transitions ($\tau_0$) and $1\rightarrow 0$ transitions ($\tau_1$) (black) with exponential fits (colour) overlaid. (i,j) Power spectral density (PSD) of the $\Delta f$ and $F_d$ time traces, respectively (colour) and the base instrument noise (grey). Lorentzian functions are overlaid, and the corner frequency ($f_C$) is indicated. (k) Correlation matrix of the $\Delta f$ and $F_d$ signals, showing the counts of a particular state (e.g. $N_{01}$ is the number of sampled points where $\Delta f$ is in state 0 and $F_d$ is in state 1).}
    \label{fig:NoiseAnalysis}
\end{figure*}

The slow ($\mathrm{Hz-kHz}$) noise at $T_b$ and $T_c$ in Figures~\ref{fig:RingofNoise} and \ref{fig:NoisySpectra} shows two-state fluctuations. The two states are here defined with the generic labels $0$ and $1$. The fact that $F_d$ fluctuates at these traps indicates that the $0$ and $1$ states have different characteristic mobilities. The total mobility $\mu_{tot}$ near the Si/SiO\textsubscript{2} interface is\cite{Sze2007}:
\begin{equation}\label{eq:mobility}
    \frac{1}{\mu_{tot}}= \frac{1}{\mu_{bulk}}+\frac{1}{\mu_{interface}}...\left(+\frac{1}{\mu_{trap}}\right)
\end{equation}
\noindent where $\mu_{bulk}$ is the lattice-limited mobility, $\mu_{interface}$ is the mobility which is limited by scattering at the Si/SiO\textsubscript{2} interface, and $\mu_{trap}$ is the mobility limited by scattering at trap sites. In the background ($B_0$), there is no trap scattering, and the final term in Equation~\ref{eq:mobility} equals zero. At static traps ($T_a$), $\mu_{trap}>0$ due to trap scattering. At fluctuating traps ($T_b$ and $T_c$), the final term in Equation~\ref{eq:mobility} fluctuates between zero and nonzero. The $0$ state, then, corresponds to the background condition, where the fast ($\mathrm{MHz}$) charge transfer between the interface trap and the silicon (and the associated energy dissipation) does not occur and the driving force is at the minimum value measured at that bias. The $1$ state corresponds to the increased loss condition, where fast charge transfer (and the associated energy dissipation) does occur. The RTN in Figures~\ref{fig:RingofNoise} and \ref{fig:NoisySpectra} is not due to tunneling between the tip (gate) and the trap, since RTN is observed even when the tip-sample separation is increased by $10~\mathrm{nm}$ (see Supplemental Materials S.1).

\subsection*{RTS characterization}

$\Delta f$ and $F_d$ both exhibit slow two-state fluctuations. While only the $F_d$ fluctuations are resolved in Figure~\ref{fig:NoisySpectra} (see Supplemental Material S.1), fluctuations in both $\Delta f$ and $F_d$ can be seen in Figure~\ref{fig:NoiseAnalysis}. Figure~\ref{fig:NoiseAnalysis} shows a RTS measured by positioning the fm-AFM tip above a fast ($T_c$-type) trap and recording $\Delta f$ and $F_d$ as a function of time. In this measurement, $V_g=-6.5~\mathrm{V}$. $\Delta(\Delta f)$ and $\Delta F_d$ show the change in $\Delta f$ and $F_d$ with respect to state 0. 

The two-state fluctuations (i.e. RTS) measured at this trap can be generically described as the reversible reactions:
\begin{subequations}\label{eq:GenericRTS}
\begin{align}
    0 &\xrightarrow{k_{01}} 1 \\
    1 &\xrightarrow{k_{10}} 0
\end{align}
\end{subequations}
\noindent with rates $k_{01}$ and $k_{10}$. The RTS amplitude ($A_{01}$) is found by fitting Gaussian distribution functions to the RTS time trace (Figure~\ref{fig:NoiseAnalysis}a-b,e-f) such that:
\begin{equation}\label{eq:RTSamplitude}
    A_{01}=|\bar{0}-\bar{1}|
\end{equation}
\noindent where $\bar{0}$ $(\bar{1})$ is the mean value of 0 (1) peak. The time between transitions is exponentially distributed (Figure~\ref{fig:NoiseAnalysis}c-d,g-h), such that the probability of a transition from the 0 to 1 (1 to 0) state ($P_{0\rightarrow 1({1\rightarrow 0})}$) is:
\begin{subequations}\label{eq:RTSexponential}
\begin{align}
    P_{0\rightarrow 1} &=\frac{1}{\bar{\tau}_0}\exp\left(-\frac{\tau_0}{\bar{\tau}_0}\right)\\
    P_{1\rightarrow 0} &=\frac{1}{\bar{\tau}_1}\exp\left(-\frac{\tau_1}{\bar{\tau}_1}\right)
\end{align}
\end{subequations}
\noindent where $\tau_{0(1)}$ is the wait time in the $0$ ($1$) state before a $0\rightarrow 1$ ($1\rightarrow 0$) transition, and $\bar{\tau}_{0(1)}$ is the average wait time in the $0$ ($1$) state. (i.e. ${k_{01}=1/\bar{\tau}_0}$ and ${k_{10}=1/\bar{\tau}_1}$ are the transition rates). The power spectral density (PSD, Figure~\ref{fig:NoiseAnalysis}i,j) of the RTS is Lorentzian in the form\cite{Machlup1954,Kirton1989}: 
\begin{equation}\label{eq:PSD}
    \mathrm{S}(f_F) = \frac{4A_{01}^2}{\left(\bar{\tau}_0+\bar{\tau}_1\right)\times\left[\left(\frac{1}{\bar{\tau}_0}+\frac{1}{\bar{\tau}_1}\right)^2+\left(2\pi f_F\right)^2\right]}
\end{equation}
\noindent where $f_F$ is the Fourier frequency. Equation~\ref{eq:PSD} shows that ${S(f_F) \propto f_F^\alpha}$, where the exponent (i.e. slope of the PSD on logarithmic axes) $\alpha=0$ for ${f_F<f_C}$ and $\alpha=-2$ for ${f_F>f_C}$, where the corner frequency $f_C$ is: 
\begin{equation}\label{eq:CornerFrequency}
    f_C=\frac{1}{\bar{\tau}_0+\bar{\tau}_1}
\end{equation}
\noindent The Lorentzian curves and $f_C$ values shown in Figure~\ref{fig:NoiseAnalysis}i-j correspond to Equations~\ref{eq:PSD} and \ref{eq:CornerFrequency} and use no fit parameters. This validates the use of Equations~\ref{eq:RTSamplitude} and \ref{eq:RTSexponential} to extract RTS amplitudes and rates.

\subsection*{Correlated fluctuations} 

The correlation between the $\Delta f$ and $F_d$ fluctuations is quantified as the phi coefficient:
\begin{equation}
    \Phi=\frac{N_{11}N_{00}-N_{10}N_{01}}{\sqrt{(N_{11}+N_{10})(N_{10}+N_{00})(N_{00}+N_{01})(N_{01}+N_{11})}}
\end{equation}
\noindent where $\Phi=1$ indicates perfect correlation, $\Phi=-1$ indicates perfect anti-correlation, and $\Phi=0$ indicates no correlation. $N_{11}$, $N_{10}$, $N_{01}$, and $N_{00}$ (Figure~\ref{fig:NoiseAnalysis}k) are the number of instances (over the time trace) where $\Delta f$ and $F_d$, respectively, are in the given state. (For example, $N_{01}$ is the number of sampled points where $\Delta f$ is in state $0$ and $F_d$ is in state $1$.) The system is defined as being in a given state based on whether the measurement falls within the full-width half-maximum of the Gaussian peak, see the Supplemental Materials. For the RTS shown in Figure~\ref{fig:NoiseAnalysis}, $\Phi=0.87$, which indicates that $\Delta f$ and $F_d$ are strongly correlated. (The departure from $\Phi=1$ is attributed to the uncertainty in assigning the 0 and 1 states to each sampled point, see Supplemental Material S.2.) This result indicates that as the interface trap switches between the $0$ and $1$ states according to Equation~\ref{eq:GenericRTS}, the interfacial capacitance and dielectric loss at the interface trap site both fluctuate. This means that there are correlated carrier number and mobility fluctuations near traps at the Si/SiO\textsubscript{2} interface.

\subsection*{Bias dependence} 

\begin{figure}[!t]
    \centering
    \includegraphics[width=\linewidth]{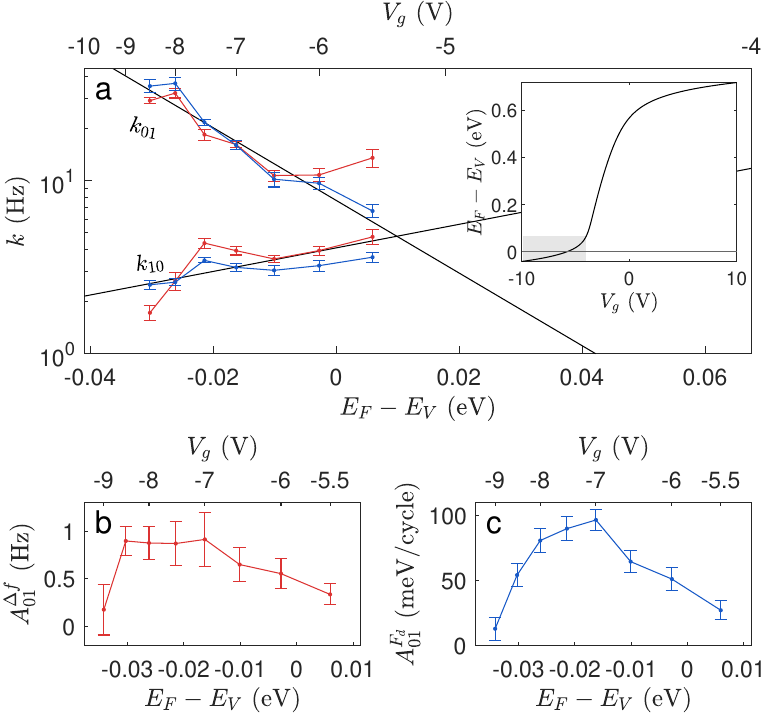}
     \caption{\textbf{Random telegraph signal (RTS) amplitude and rate dependence on applied bias.} fm-AFM frequency shift ($\Delta f$, red) and drive ($F_d$, blue) RTS rates (a) and amplitudes (b,c) measured simultaneously at a single fast-switching ($T_c$-type) trap site at variable bias $V_g$. The inset in (a) shows the modelled non-linear relationship between $V_g$ and $E_F-E_v$ (the difference between the Fermi level and the valence band edge). The shaded region of the inset shows the x-axis range of (a), where notably the top $x$-axis is non-linear. The $E_F-E_V=0$ line is shown.}
    \label{fig:RatesAmplitudes}
\end{figure}

Figure~\ref{fig:RatesAmplitudes} shows the bias-dependent RTS rates and amplitudes of an isolated trap at the Si/SiO\textsubscript{2} interface. Figure~\ref{fig:RatesAmplitudes}a shows that for both the $0\rightarrow1$ and $1\rightarrow0$ transitions, $\log(k)$ varies linearly with $E_F-E_V$, where $E_V$ is the energy of the valence band edge. The trap energy $E_T$ is fixed with respect to the band edges, meaning that as the surface potential varies with $V_g$, $E_T$ varies with respect to the Fermi level $E_F$. The logarithmic trends in Figure~\ref{fig:RatesAmplitudes}a can be attributed to thermal activation of the reactions in Equation~\ref{eq:GenericRTS}\cite{Costanzi2017,Brower1990} according to:
\begin{equation}\label{eq:Arrhenius}
    k = k^\mathrm{o}\exp\left(-\frac{E_{A}(V_g)}{k_BT}\right)
\end{equation}
\noindent where $k$ is the RTS rate ($k_{01}$ or $k_{10}$), $k^\mathrm{o}$ is the attempt frequency ($k_{01}^\mathrm{o}$ or $k_{10}^\mathrm{o}$), $E_A$ is the bias ($V_g$)-dependent activation energy ($E_{01}$ or $E_{10}$), and $T$ is the temperature.  

Note that there is a nonlinear relationship between $V_g$ and $E_F-E_V$ due to the nonlinear bias-dependent surface potential. $k$, correspondingly, is not expected to vary exponentially with $V_g$, which is why the top $x$-axes in Figure~\ref{fig:RatesAmplitudes} are non-linear. In this work, the relationship between the experimental $V_g$ and $E_F-E_V$ was found by modelling the tip-sample junction as a metal-insulator-semiconductor capacitor (see \cite{Cowie2023} for details). The modelled relation when the cantilever is at the bottom of its oscillation -- where the tip-sample force contribution is largest -- is shown in the Figure~\ref{fig:RatesAmplitudes}a inset. 

The RTS amplitudes of the isolated trap are also bias-dependent, as shown in Figure~\ref{fig:RatesAmplitudes}b-c. The $\Delta f$ fluctuation amplitude ($A_{01}^{\Delta f}$) and $F_d$ fluctuation amplitude ($A_{01}^{F_d}$) both appear to peak where $V_g$ is between $V_C^{bot}$ and $V_C^{top}$ ($\sim -7~\mathrm{V}$, recall Figure~\ref{fig:NoisySpectra}b). At this peak bias, loss is maximized when the trap is in the $1$ state, since the trap occupancy has the highest probability of switching over every cantilever oscillation cycle. In the $0$ state, no charge switching occurs, so the difference in loss between the $0$ and $1$ states is maximized, and $A_{01}^{F_d}$ peaks. The $A_{01}^{\Delta f}$ peak indicates that the change in the interfacial capacitance associated with states $0$ and $1$ is also largest between the crossing points.

\subsection*{Emergent $1/f$ trend}

\begin{figure}[!t]
    \centering
    \includegraphics[width=\linewidth]{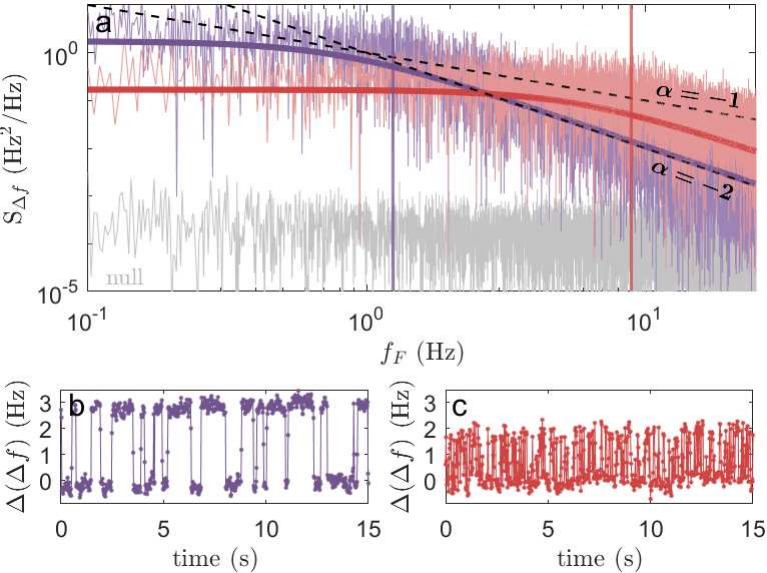}
     \caption{\textbf{Emergent 1/f trend for an ensemble of random telegraph signals.} (a) Power spectra ($S$) as a function of the Fourier frequency ($f_F$) of the fm-AFM frequency shift ($\Delta f$) channel for two traps: A slow-switching $T_b$-type trap, purple; and a fast-switching $T_c$-type trap, red); as well as the background ($B_0$, grey). Lorentzian fits and corner frequencies are shown for both traps in their respective colours. The black lines show $S \propto f_F^\alpha$ for $\alpha=-1$ and $-2$ (indicated). (b-c) Time traces of both traps in their respective colours. The time traces were each measured for $500~\mathrm{s}$; only the first $30~\mathrm{s}$ are shown.}
    \label{fig:AddingLorentzians}
\end{figure}

Figure~\ref{fig:AddingLorentzians} shows the PSDs and RTS time traces of two traps measured at the Si/SiO\textsubscript{2} interface. Individually, each trap is Lorentzian according to Equation~\ref{eq:PSD}, with an $\alpha=-2$ trend above its corner frequency. In the region between their corner frequencies (${f_F\sim 1-10~\mathrm{Hz}}$), $\alpha\approx-1$. In this electrostatic force microscopy methodology, traps are measured individually, and so exhibit Lorentizian power spectra. In a MOSFET device, however, depending on its size and quality, many traps can contribute to the total noise spectrum. The $\alpha=-1$ dependence of MOSFET PSDs is commonly attributed to an ensemble of two-state fluctuators with a range of amplitudes and corner frequencies\cite{Costanzi2017,Mueller1994,vanderWel2007}. 

Considering the emergence of a $1/f$ trend in the noise, note that Figure~\ref{fig:RingofNoise} demonstrates that these interface traps, even when located mere nanometers apart, exhibit a variety of RTS timescales. Furthermore, the ring positions in Figure~\ref{fig:RingofNoise} were stable over several months of measurements at
large positive and negative bias (${V_g=-10:10~\mathrm{V}}$) at room temperature. This indicates that the RTS reactions ($0\rightarrow 1$ and $1\rightarrow 0$) are permanently more favorable at certain locations, which, given Equation~\ref{eq:Arrhenius}, can be attributed to a lowering of the RTS reaction activation energies $E_{01}$ and $E_{10}$ by a permanent nanoscale variability in the local electrostatic landscape. In the Si/SiO2 interfacial system studied here, nanoscale spatial heterogeneities in the amorphous SiO\textsubscript{2} likely modify the electrostatic landscape near traps, giving rise to the observed variability in RTS rates that collectively can produce an emergent $1/f$ trend as illustrated in Figure~\ref{fig:AddingLorentzians}.

\section*{Discussion}\label{Sec:Discussion}
\begin{figure}[!b]
    \centering
    \vspace{-4mm}
    \includegraphics[width=0.95\linewidth]{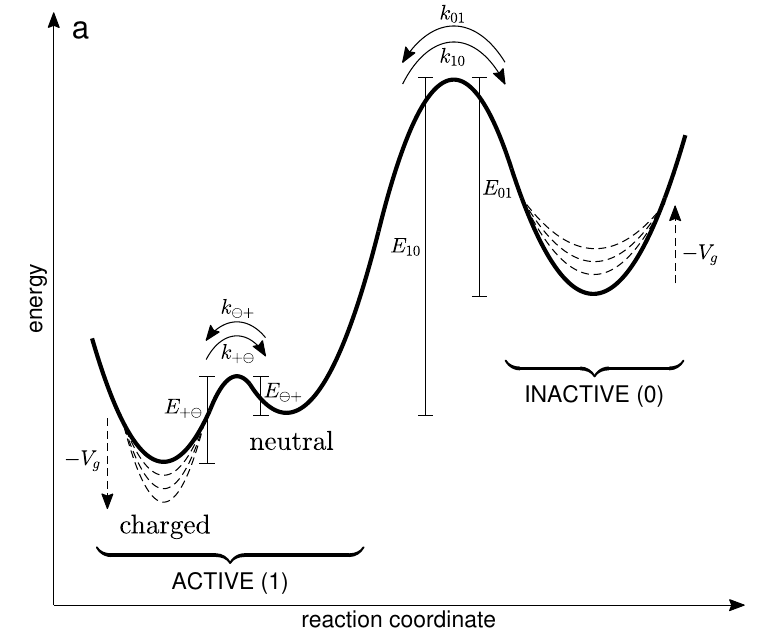}
    \includegraphics[width=0.95\linewidth]{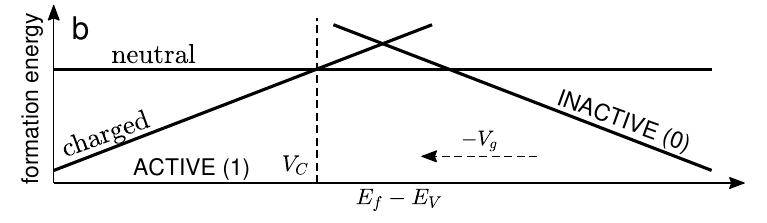}
     \caption{\textbf{Four-fold rate competition reaction schematic}. (a) Reaction diagram for trap activation and inactivation ($k_{01}$ and $k_{10}$) and charging and discharging ($k_{\mathrel{\ominus}{+}}$ and ${k_{\mathrel{+}{\ominus}}}$). The dashed lines show that with increasing negative bias ($V_g$), $E_{01}$ decreases and $E_{\mathrel{+}{\ominus}}$ increases. (b)~Formation energy diagram for the active state (which can be neutral or charged) and the inactive state. The dashed arrow indicates the direction of increasing negative bias. The crossing point $V_C$, at which the trap level equals the Fermi level, is shown as a vertical dashed line.}
    \label{fig:ThermalReaction}
\end{figure}

The fm-AFM tip-sample junction resembles a metal-insulator-semiconductor capacitor. The two-state fluctuations measured in this work, therefore, may also contribute to gate noise in MOSFET devices, which, like the noise measured above, exhibit low characteristic frequencies on the order of $\mathrm{Hz}$\cite{Kirton1989,Kirton1989b,Grasser2020}. In MOSFET devices, RTN is commonly attributed to tunneling between an oxide trap and the channel, where the tunneling rate is determined by the position of the trap within the oxide. Carrier number fluctuations are attributed to capture and emission of charge from the channel by the trap, and mobility fluctuations are attributed to variable scattering from the trap, since the scattering cross section depends on the trap charge state. However, this common description of the RTN mechanism in MOSFET devices has two significant shortcomings. First, RTN frequencies increase with temperature above $10~\mathrm{K}$, indicating thermally activated carrier exchange (Equation~\ref{eq:Arrhenius}) rather than tunneling\cite{Fleetwood2023}. Second, predicted Coulomb scattering from oxide traps is very small, because oxide traps are spatially removed from the channel\cite{Fleetwood2023}. Alternative RTN mechanisms should therefore be considered. A mechanism involving interface traps will be discussed below.

The bias spectral peaks shown in Figure~\ref{fig:NoisySpectra}b correspond to charge state switching of donor-like interface traps\cite{Cowie2023}, which exhibit fast charging and discharging rates ($k_{\mathrel{\ominus}{+}}$ and $k_{\mathrel{+}{\ominus}}$, respectively) on the order of $\mathrm{MHz}$\cite{Labidi2015,Cowie2023}. These charging/discharging timescales are much faster than slow $\mathrm{Hz}-\mathrm{kHz}$ RTS switching between states $0$ and $1$ ($k_{01}$ and $k_{10}$). The measurements shown in this work, therefore, point to a four-fold competition of rates, as illustrated in Figure~\ref{fig:ThermalReaction}a. The $0$ state is an `inactive' state, where the trap occupancy is constant. The $1$ state is `active', meaning that the donor-like interface trap occupancy can switch between neutral and charged. Switching between the active ($1$) and inactive ($0$) states is slow ($\mathrm{ms-s}$), and trap charging and discharging when the trap is in the active state is fast ($\mathrm{ns}$). In other words, in the active ($1$) state, the trap is able to be charged and discharged, and energy is dissipated by phonon scattering each time the trap occupancy switches\cite{Cowie2023}. The inactive ($0$) state, however, does not experience this fast charge switching and associated energy loss. In Figure~\ref{fig:ThermalReaction}b, the formation energies for the 0 and 1 states are shown as linearly dependent on $E_F-E_V$, given Equation~\ref{eq:Arrhenius} and the linear RTS rates shown in Figure~\ref{fig:RatesAmplitudes}a.

\subsection*{Interface trap hydrogen passivation mechanism} 

A possible noise mechanism, to explain the results presented here, including the four-fold competition of rates illustrated in Figure~\ref{fig:ThermalReaction}, is the slow spontaneous passivation and de-passivation of a Pb center, i.e. an interface trap dangling bond (DB)\cite{Fleetwood2023,Stirling2000,Goes2018,Herring2001} with hydrogen. In the passivated state (${\mathrm{Pb}\textrm{-}\mathrm{H}}$, i.e. ${\mathrm{Si_3\equiv Si}\textrm{-}\mathrm{H}}$), the trap would be inactive ($0$), but in the de-passivated state ($\mathrm{Pb\cdot}$, i.e. ${\mathrm{Si_3\equiv Si\cdot}}$), the trap would be active ($1$), meaning that its charge state could rapidly switch between neutral and charged, due to the proximity of the Pb center to the silicon surface charge density. 

While details of individual Pb passivation at the Si/SiO\textsubscript{2} interface are difficult to ascertain directly, the simpler and related system of hydrogen terminated Si(001) is well-studied, and scanning tunneling microscopy (STM) measurements provide atomic resolution details of DB passivation\cite{Pitters2024,Taucer2014}. In particular, bi-stabilities attributed to hydrogen diffusion at a DB site at the bare Si(001) surface have been measured at room temperature on the order of seconds\cite{Bellec2010,Stokbro2000}. These timescales are very similar to the long $0$ and $1$ state lifetimes shown in Figures~\ref{fig:NoiseAnalysis} and \ref{fig:RatesAmplitudes}a, and thus support the hydrogen passivation of Pb centres as a possible explanation for the slow noise measured in this work, at the Si/SiO\textsubscript{2} interface. 

In this proposed slow noise mechanism, the Pb center can be passivated (inactivated) by reacting with either molecular\cite{Stesmans2000, Brower1990,Ragnarsson2000} or atomic\cite{Cartier1993,Ragnarsson2000} hydrogen according to:
\begin{subequations}\label{eq:Passivation}
\begin{align}
    \mathrm{Pb\cdot} + \mathrm{H\textsubscript{2}}&\xrightarrow{~~} \mathrm{Pb}\textrm{-}\mathrm{H}+\mathrm{H\cdot}\\
    \mathrm{Pb\cdot} + \mathrm{H\cdot}&\xrightarrow{~~} \mathrm{Pb}\textrm{-}\mathrm{H}
\end{align}
\end{subequations}
\noindent The Pb center can be de-passivated (activated) according to\cite{Cartier1993,Tsetseris2004,Stesmans2000,Brower1990}:
\begin{subequations}\label{eq:Activation}
\begin{align}
    \mathrm{Pb}\textrm{-}\mathrm{H} + \mathrm{H\cdot} &\xrightarrow{~~} \mathrm{Pb\cdot} + \mathrm{H}_2\\
    \mathrm{Pb}\textrm{-}\mathrm{H} &\xrightarrow{~~} \mathrm{Pb\cdot} + \mathrm{H\cdot}
\end{align}
\end{subequations}

\noindent 

The DB state in Equations~\ref{eq:Passivation} and \ref{eq:Activation} is neutral (i.e. $\mathrm{Pb}\cdot$ is equivalent to $\mathrm{Pb}^\ominus$), but a donor-like DB can be positively charged according to: 
\begin{subequations}\label{eq:ChargingDischarging}
\begin{align}
    \mathrm{Pb}^\ominus+\mathrm{h}^+ &\xrightarrow{k_{\mathrel{\ominus}{+}}} \mathrm{Pb}^+~~~~~~~~~~~\mathrm{(charging)}\\
    \mathrm{Pb}^+ &\xrightarrow{k_{\mathrel{+}{\ominus}}} \mathrm{Pb}^\ominus+\mathrm{h}^+~~~~\mathrm{(discharging)}
\end{align}
\end{subequations}
\noindent where $k_{\mathrel{\ominus}{+}}$ and ${k_{\mathrel{+}{\ominus}}}$ are the charging and discharging rates, respectively. 

If this mechanism is indeed responsible for the observed RTN, then there should be a plausible explanation for passivation/depassivation of the Pb defects  at room temperature, while exposed to a time varying ($\mathrm{kHz}$) gate (tip) bias, of order $1-10~\mathrm{V}$. In STM experiments on hydrogen terminated silicon, DB activation occurs when tunneling electrons transfer enough energy to the Si-H bond to cleave it. This tunnelling electron stimulated hydrogen desorption can be reliably controlled and is exploited to perform atomically precise lithography\cite{Shen1995}. The bond cleaving mechanisms at work here are well understood. At high sample biases ($>\sim6.5~\mathrm{V}$), a single electron process is accessed, in which direct electronic excitation of the Si-H bond from the bonding to anti-bonding state serves to remove the hydrogen\cite{Hersam2002}. At lower sample tunnel biases ($<\sim 6.5~\mathrm{V}$) Si-H vibrational modes  are excited by inelastic scattering of the tunneling electrons with the Si-H bond\cite{Bellec2010,Stokbro2000,Shen1995,Moller2017,Pavliek2017,Persson1988}; by pumping these vibrational modes, in a multi-electron process, the bond can be broken.  Similar mechanisms are also understood to occur in hot carrier induced damage in CMOS devices\cite{Jech2021}. In this case, sufficiently energetic electrons in the channel current interact with and break Si-H bonds at the gate oxide interface creating charge traps and degrading device performance. In our fm-AFM experiment, we suggest that DB activation at the Si/SiO\textsubscript{2} interface can occur when Si-H bonds are excited sufficiently to overcome barrier $E_{01}$ in Figure~\ref{fig:ThermalReaction}, either directly or indirectly, by the oscillating electric field introduced by the biased AFM tip. As the AFM tip oscillates, bending the silicon bands at the sample surface, the interfacial carrier density continually re-organizes, introducing a local current density at the Si/SiO\textsubscript{2} interface. In analogy with the STM induced hydrogen desorption from Si(001) and hot carrier damage in CMOS devices, this surface charge reorganization current density and/or the associated electric field\cite{Jech2021} may be responsible for cleaving of select Si-H bonds at the Si/SiO\textsubscript{2} interface in the sample measured here using fm-AFM.

DB passivation, on the other hand, can occur when the DB is exposed to interstitial atomic and/or molecular hydrogen\cite{Cartier1993,Tsetseris2004}, both of which can diffuse near the Si/SiO2 interface\cite{Stesmans2000}. In MOSFET devices, hydrogen is intentionally introduced to passivate the oxide interface traps and improve device performance. The sample measured in this work, however, was not treated with hydrogen post-oxide growth to passivate DBs at the interface.  Rather, the sample  was fabricated using the same procedure as for hydrogen resist lithography, where the atomically flat Si(001) surface was passivated in UHV with atomic hydrogen and subsequently capped by $3~\mathrm{nm}$ of epitaxial silicon\cite{Stock2020}; when exposed to atmosphere, $\sim 1~\mathrm{nm}$ of native SiO\textsubscript{2} is formed at the surface. Due to this process we expect only a fraction of interfacial DBs to be passivated. In the Pb passivation (slow fluctuation) mechanism proposed here, hydrogen present as interstitials or bound at various defects including other Pb centres may supply a reservoir available to passivate the Pb centres that have been activated under the influence of the oscillating tip bias. The concentration and diffusivity of the interstitial hydrogen will therefore contribute to the DB passivation rate $k_{10}$\cite{Stesmans2000}. Furthermore, the passivation rate of the PB centre will depend on its charge state\cite{Ragnarsson2000}; this is consistent with the decrease of $k_{10}$ with increasing negative bias measured here (Figure~\ref{fig:RatesAmplitudes}a).

\section*{Conclusions}\label{Sec:Conclusions}
We investigated the nanoscale spatial heterogeneity of random telegraph noise at the Si/SiO\textsubscript{2} interface. Our measurements show that two-state fluctuations are localized at individual interface charge traps. The traps measured here exhibit a range of characteristic RTS timescales, even though the traps were located mere tens of nanometers apart, and positioned equidistant from the gate (fm-AFM tip). This indicates that the RTS rate is affected by nanoscale heterogeneities in the electrostatic landscape, which is here attributed to the heterogeneity of the amorphous SiO\textsubscript{2} overlayer. The range of RTN rates and corner frequencies that results from this inherent spatial heterogeneity gives rise to a collective $1/f$ power spectrum, which is a ubiquitous phenomenon in silicon-based semiconducting devices.

The nanoscale spatial heterogeneity in the noise landscape measured here implies significant variability in the noise characteristics of nanoscale silicon field-effect devices. Furthermore, these findings are of particular importance for advancing silicon-based quantum sensors and computers, where qubits are spaced roughly $1~\mathrm{nm}$ apart within a silicon lattice. In such devices, each qubit interacts with its local silicon electronic bath. Thus, the nanoscale electronic heterogeneities measured in this work indicate that each qubit could be susceptible to a unique noise landscape, potentially leading to distinct noise profiles for individual qubits. 

The fm-AFM measurement methodology used in this work provides a unique opportunity to simultaneously measure two drastically different noise timescales ($\sim \mathrm{ns}$ and $\sim \mathrm{s}$) associated with individual traps. The fast rates, $k_{\mathrel{\ominus}{+}}$ and ${k_{\mathrel{+}{\ominus}}}$, are associated with energy loss, and introduce a phase lag between the $\sim 300~\mathrm{kHz}$ bias modulation (in this case, applied by the oscillating top gate i.e. fm-AFM tip) and the sample response. The slow rates, $k_{01}$ and $k_{10}$, are here fully resolved RTSs. Together, these fast and slow processes imply a four-fold competition of rates, where one RTS state ($1$) is associated with appreciable energy loss, and the other ($0$) is not. These findings address long-standing questions about the behaviour of traps at the Si/SiO\textsubscript{2} interface. 




\matmethods{
\subsection*{fm-AFM}\label{sec:fm-AFM}
The fm-AFM cantilever is maintained at a constant oscillation amplitude $A$ on its resonance frequency $\omega$ using a self-excitation loop which applies a periodic driving force with amplitude $F_d$. In the linear force-distance regime, the tip-sample separation varies sinusoidally according to ${z_{ins}=A\cos(\omega t)}$. The components of $F_{ts}(t)$ which are in-phase with $z_{ins}(t)$ lead to a shift in the cantilever resonance frequency (${\Delta\omega=2\pi\Delta f}$) with respect to its free natural resonance $\omega_o$, and the out-of-phase force components lead to an increase in $F_d$ according to\cite{Holscher2001,Sader2005,Kantorovich2004}:
\begin{subequations}\label{eq:dfdg}
\begin{align}
    \Delta\omega=\omega-\omega_o = \frac{-\omega_o}{2 kA}\frac{\omega_o}{\pi}\int_{0}^{2\pi/\omega}\partial t~F_{ts}(t)\cos(\omega t)\\
    F_d = \frac{kA}{Q}-\frac{\omega_o}{\pi}\int_{0}^{2\pi/\omega}\partial t~F_{ts}(t)\sin(\omega t)
\end{align}
\end{subequations}
\noindent where $k$ and $Q$ are the cantilever spring constant, and quality factor, respectively. 

Here, $F_{ts}$ is predominantly electrostatic, and is related to the charge density at the silicon surface\cite{Hudlet1995}. $\Delta f$ is therefore a measure of the interfacial capacitance. The first term in Equation~\ref{eq:dfdg}b represents intrinsic losses due to cantilever damping; the second term accounts for additional energy losses in the sample. Here, $\omega\approx 300~\mathrm{kHz}$, which is in a low-frequency regime where scattering of mobile carriers at the surface dominates energy loss, such that the loss tangent $\tan(\delta)$ is:
\begin{equation}
    \tan(\delta)\approx\frac{\sigma}{\omega\epsilon\epsilon_o}
\end{equation}
\noindent where $\sigma$ is the conductivity and $\epsilon\epsilon_o$ is the permittivity. In other words, $F_d$ is a measure of the equivalent series resistance of the interfacial capacitance.

\subsection*{Experimental setup}
All results were measured at room temperature ($\sim 300~\mathrm{K}$) in UHV ($\sim 10\mathrm{e}{-10}~\mathrm{mbar}$) using a JEOL JSPM-4500A system. A beam deflection detection mechanism was used for fm-AFM, with Nanosensors platinum-iridium
coated silicon cantilevers (PPP-NCHPt, ${f_o\approx300~\mathrm{kHz}}$, ${k\approx42~\mathrm{N/m}}$, ${Q\approx18000}$, and ${A\approx 6~\mathrm{nm}}$), and a Nanonis controller with a $20~\mathrm{ms}$ sampling frequency. Bias spectra (Figure~\ref{fig:NoisySpectra}) were measured over $\sim 30~\mathrm{s}$, and show the positive-negative and negative-positive sweeps superimposed. For multipass images (Figure~\ref{fig:RingofNoise}) $\Delta f=-3~\mathrm{Hz}$ and $V_g=0~\mathrm{V}$ for the first pass, and $V_g$ was set to the displayed value following the first-pass topography for subsequent passes. For time trace measurements (e.g. Figure~\ref{fig:NoiseAnalysis}a,e), the $z$-controller was turned off. Vertical ($z$) drift was $<1~\mathrm{nm}$ for every measured time trace, and the tip was re-approached before and after every time trace. Lateral ($x,y$) drift was estimated to be $<2~\mathrm{nm}$. The time traces corresponding to the data points in Figure~\ref{fig:RatesAmplitudes} were measured out of order, so the bias-dependent trends presented in this work cannot be attributed to spatial drift. 

\subsection*{Sample fabrication}
The $300~\mathrm{\mu m}$ thick n-type silicon sample was phosphorous-doped, with donor concentration $9.0\times 10^{14}/\mathrm{cm^3}$. The atomically clean Si(001) surface was passivated by atomic hydrogen in UHV at $250~\mathrm{C}$, and then capped by $3~\mathrm{nm}$ of epitaxial intrinsic Si at $250 \mathrm{C}$ following a 10-monolayer room temperature locking layer and a $15~\mathrm{s}$ $250~\mathrm{C}$ rapid thermal anneal\cite{Stock2020}. Upon exposure to atmosphere, $1~\mathrm{nm}$ of self-limited native SiO\textsubscript{2} formed at the surface. 

\subsection*{Model parameters} The MIS model\cite{Sze2007,Hudlet1995,Cowie2022,Cowie2023} parameters for the inset in Figure~\ref{fig:RatesAmplitudes}a are: Closest $z_{ins}=12~\mathrm{nm}$, tip radius $5~\mathrm{nm}$, $\epsilon=11.7$,
electron affinity $4.05~\mathrm{eV}$, tip work function $4.75~\mathrm{eV}$, electron and hole effective masses $1.08$ and $0.56$, n-type
dopant density $5\times 10^{17}\mathrm{/cm^3}$, and band gap $0.7~\mathrm{eV}$ (which can be attributed to surface band gap narrowing due to the large interface trap density, as in \cite{vanVliet1986,King2010}).}
\showmatmethods{}

\acknow{This research was supported by Natural Sciences and Engineering Research Council of Canada (NSERC) Alliance Grants -- Canada-UK Quantum Technologies, an NSERC Discovery Grant, and Fonds de recherche du Qu\'{e}bec -- Nature et technologies, as well as the Engineering and Physical Sciences Research Council [grants EP/R034540/1, EP/V027700/1, and EP/W000520/1] and Innovate UK [grant 75574]. For the purpose of open access, the author has applied a Creative Commons Attribution (CC BY) licence to any Author Accepted Manuscript version arising.}
\showacknow{}

\bibsplit[7]

\bibliography{Research_references}

\end{document}


\begin{center}
    \Large Supplemental material
    \\
    \normalsize for
    \\
    \large \textbf{Spatially resolved random telegraph fluctuations \\ of a single trap at the Si/SiO\textsubscript{2} interface}
    \\ \vspace{5mm}
    \normalsize Megan Cowie, Taylor T.Z. Stock, Procopios C. Constantinou, Neil J. Curson, and Peter Gr\"{u}tter
\end{center}

\vspace{10mm}
\large \noindent \textbf{S.1: Distance dependence of donor-like defect noise} \normalsize

The slow ($\mathrm{ms-s}$) two-state noise measured above donor-like traps cannot be attributed to tunneling between the tip and sample, since noise with similar timescales is measured even when the tip-sample separation is increased (i.e. ``tip lifted") by several nanometers, as shown in Figure~S1. (Note that the noise amplitude in the $\Delta f$ channel is very small as compared to the background -- i.e. $\sim (1~\mathrm{Hz})/(500~\mathrm{Hz})\approx 0.2\%$, which is why it is indistinguishable in Figure~S1a. In the $F_d$ channel (Figure~S1b), however, the noise amplitude is comparably large -- i.e. $\sim (80~\mathrm{meV/cycle})/(\mathrm{300~meV/cycle})\approx 25\%$.) If the slow noise were due to tunneling between the tip and sample, the noise timescale would be expected to increase significantly as tip lift increased. 

\begin{figure}[!b]
\centering    
\includegraphics[width=\linewidth]{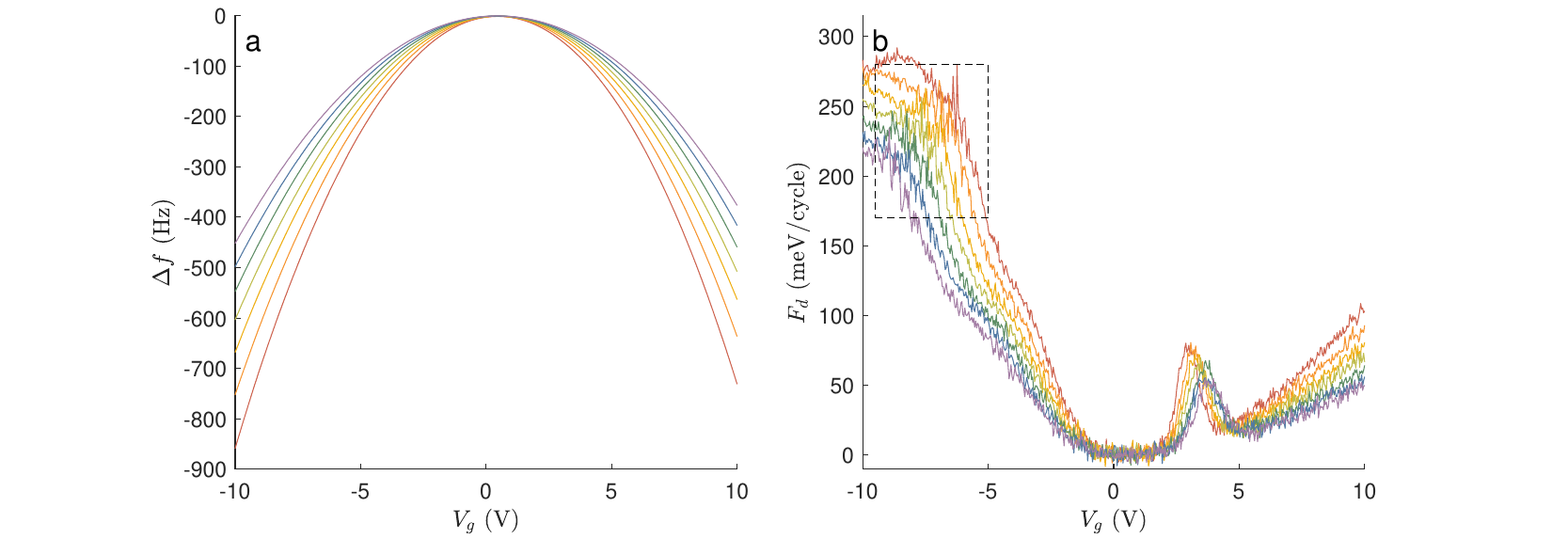}
\caption*{\textbf{Fig.~S1. Distance-dependent bias spectra.} Bias spectra near a donor-like trap (seen at negative bias) and an acceptor-like trap (seen at positive bias) at variable tip lift (red $=0~\mathrm{nm}$, purple $=9~\mathrm{nm}$, in steps of $1~\mathrm{nm}$). The dashed rectangle indicates noise due to the donor-like trap. (The acceptor-like trap does not exhibit additional noise compared to the background.)}
\end{figure}

\vspace{10mm}
\large \noindent \textbf{S.2: Two-state noise analysis} \normalsize

The methodology to assign states $0$ and $1$ for an RTS is demonstrated in Figure~S2. First, the signal is subtracted from its running mean (with an averaging timescale much longer than the RTS noise timescale). $\Delta(\Delta f)$ and $\Delta F_d$ are the difference between $\Delta f$ or $F_d$, respectively, and their running means. The subtracted signals are offset so that the $0$ state occurs at $0$ (Figure~S1a,d). Two Gaussian distribution functions are then fitted to histograms of the $\Delta(\Delta f)$ and $\Delta F_d$ time traces (Figure~S1b,e). The shaded regions in Figure~S2a,d show the standard deviation of these Gaussian fits. 

States $0$ and $1$ are assigned by comparing the RTS time traces to the Gaussian fits. First, the first data point is identified as being in the ``up" state ($0$ for $\Delta f$ and $1$ for $F_d$) or the ``down" state. Then, each data point is compared to the previous one. The "up" state flips to "down" if the value is less than the "down" shaded region maximum, and the "down" state flips to "up" if the value is greater than the "up" shaded region minimum. The circled points (Figure~S2a,d) show these state flips. The arrows point to mismatches in the $\Delta f$ and $F_d$ flips; these appear to be state assignment errors, rather than non-correlation between the $\Delta f$ and $F_d$ RTSs. Specifically, the $F_d$ RTS amplitude is larger than the intrinsic noise (Gaussian peak width), and the flip values are approximately normally distributed in each state (Figure~S2f). The $\Delta f$ RTS amplitude is small compared to the intrinsic noise, and the flip values are not normally distributed (Figure~S2c). This indicates an overestimation of $\Delta f$ flips, and manifests as shorter $\Delta f$ RTS timescales. 

\begin{figure}[!b]
\centering    
\includegraphics[width=\linewidth]{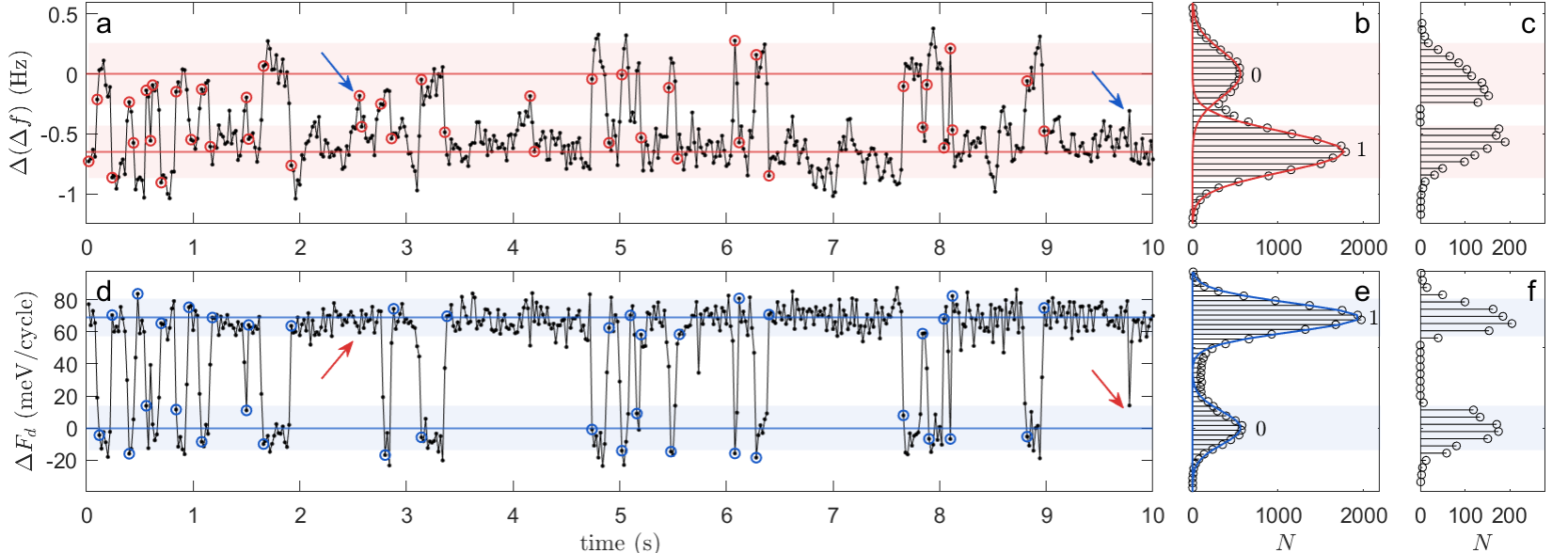}
\caption*{\textbf{Fig.~S1. RTS state assignment methodology.} RTS of a donor-like defect at the Si/SiO\textsubscript{2} interface measured at $V_g=-6.5~\mathrm{V}$. The timescales found using the methodology described above are: ${\tau_0^{\Delta f}=0.093\pm0.007~\mathrm{s}}$, $\tau_1^{\Delta f}=0.28\pm0.014~\mathrm{s}$, $\tau_0^{F_d}=0.10\pm0.01~\mathrm{s}$, $\tau_1^{F_d}=0.33\pm0.02~\mathrm{s}$.}
\end{figure}